\documentclass[twocolumn,aps,prl,preprintnumbers,amsmath,amssymb,nofootinbib,letter,floatfix]{revtex4-1}\usepackage[dvips]{graphicx}
\usepackage{color}
\usepackage{ulem}

\begin{document}

\newcommand{\BejLen}{l_{\mathrm{B}}}
\newcommand{\AArm}{\, \mathrm{\AA}}
\newcommand{\Mrm}{\, \mathrm{M}}
\newcommand{\rbf}{\mathbf{r}}
\newcommand{\kbf}{\mathbf{k}}
\newcommand{\Gk}{\widetilde{G}(\mathbf{k})}
\newcommand{\ehat}{\mathbf{\hat{e}}}
\newcommand{\kbt}{k_{\mathrm{B}}T}
\newcommand{\DebLen}{\lambda_{\mathrm{D}}}
\newcommand{\DebFreq}{\kappa_{\mathrm{D}}}
\newcommand{\captionset}{\addtolength{\leftskip}{1.1cm}\addtolength{\rightskip}{1.2cm} \small\textsf}
\newcommand{\be}{\begin{equation}}
\newcommand{\ee}{\end{equation}}
\newcommand{\bea}{\begin{eqnarray}}
\newcommand{\eea}{\end{eqnarray}}

\title{The Dielectric Constant of Ionic Solutions: \\A Field-Theory Approach}

\author{Amir Levy, David Andelman$^{*}$}
\affiliation{Raymond and Beverly Sackler School of Physics and Astronomy,
Tel Aviv University, Ramat Aviv 69978, Tel Aviv, Israel}

\author{Henri Orland}
\affiliation{Institut de Physique Th\'eorique, CE-Saclay, CEA,
F-91191 Gif-sur-Yvette Cedex, France}

\date{Jan 23, 2012}

\begin{abstract}

We study the variation of the dielectric response of a dielectric liquid (e.g. water)
when a salt is added to the solution.
Employing field-theoretical methods we expand the Gibbs free-energy to first order in
a loop expansion and calculate self-consistently the dielectric constant.
We predict analytically the dielectric decrement which depends on the ionic
strength in a complex way. Furthermore, a qualitative
description of the hydration shell is found and is
characterized by a single length scale. Our prediction fits rather well a
large range of concentrations for different salts using only one fit parameter related to
the size of ions and dipoles.

\end{abstract}

\maketitle


Electrostatic interactions in aqueous media between dipoles and charged objects
such as ions, colloidal particles and interfaces, play an important role in electrochemistry,
biology and materials science. The century-old Poisson-Boltzmann (PB) theory gives a simple
and powerful description for numerous systems, taking into account only Coulombic interactions
on a mean-field level, treating ions as point-like particles and the aqueous
solution as a continuous and homogeneous dielectric medium with a constant dielectric constant,
$\varepsilon_w$. The PB theory succeeded over the years in capturing much of the underlying
physics of electrolyte solutions, and, in particular,
is successful when applied to monovalent ions and weak surface charges~\cite{Israelachvili,andelman1,PolarMolecules,Onsager}.

The PB theory has several limitations. It does not take into account neither the correlations between
the charges, nor does it allow for any fluctuations beyond mean field. This leads to significant unaccounted
corrections in cases of high charge density, especially near charged surfaces and interfaces~\cite{ByondPB}.
In order to improve upon the PB theory, several extensions have been offered in recent years, and effects of
correlations and fluctuations for charged interfaces were considered, for example,
by using integral equation theories \cite{IntegralEqn1}. In other approaches \cite{IonSpecInt},
the PB theory was modified in a simple and elegant way to include steric and other ionic-specific effects
preventing ions from accumulating near charged surface for very high ionic concentrations.
Furthermore, in the well known Deryagin-Landau-Verwey-Overbeek (DLVO) theory~\cite{DLVO}, van der Waals
attractive interactions were added to the electrostatic repulsion in order
to explain charged colloidal stability. More recently,  Monte-Carlo (MC) simulations and
Molecular Dynamics (MD) were employed in order to study specific solvents and solutes and the
interaction between them \cite{MonteCarlo,MolecularDynamics1,MD1,MD2,MD3}.

The PB theory as well as others {\it primitive models} of ionic solutions~\cite{primitive}
assumes that the ions are immersed in a continuum dielectric background characterized
by the dielectric constant of water, $\varepsilon_w$. Hence, in order to model the experimentally
known dielectric decrement phenomenon~\cite{IonSpecInt,Experiment1,Experiment2,Experiment3},
one needs to employ more refined theories,
which take  into account ion-dipole correlations and fluctuations. The electric decrement
stems from the fact that the local electric field around each ion is greater than the
external field, and orients the dipolar water molecules in its vicinity.
This creates a {\it hydration shell} \cite{IonSpecific} that encircles the ions.
The total response of dipoles to the external field is thus smaller and leads to a
reduction in the dielectric  constant. In dilute solutions
the dielectric constant depends linearly on $c_s$, the salt concentration,
$\varepsilon(c_s) = \varepsilon_w - \gamma c_s$, where $\gamma$ is the coefficient
of the linear term. The value of $\gamma$ is ion specific and ranges from $8$\,M$^{-1}$ to $20$\,M$^{-1}$,
for concentrations up to $1.5$\,M. At higher $c_s$ values, noticeable deviations from linearity
are observed and are due to ion-ion interactions~\cite{Experiment1,Experiment2,Experiment3}.

In this Letter we go beyond the standard PB theory and investigate the microscopic origin of the
dielectric decrement. We construct the field theory for
a grand-canonical ensemble of point-like ions and dipoles. On a mean-field level,
we obtain an analytic description of the hydration shell around ions and calculate the average dielectric constant
in the shell.
Then, using a field-theoretical approach we examine
the dielectric decrement by a one-loop
calculation. A closed-form formula for the dielectric constant is
obtained and depends on a single length-scale, related to the typical size of ions and water molecules.
Our results show a substantial contribution arising from the one-loop correction terms,
and deviates from the linear relation,
$\varepsilon(c_s) = \varepsilon_w - \gamma c_s$.
We fit the calculated $\varepsilon(c_s)$ and compare it with the experimental data for several monovalent ions.
The results fit rather well for a wide range of ionic
concentrations, even in the non-linear decrement regime.

Although our formulation is very general, in this Letter we focus on the dielectric constant
variation for ionic solutions. The solvent (water) is modeled as a liquid of point-like permanent
dipoles of  dipolar moment $\mathbf{p}_0$ and the symmetric  $1{:}1$ monovalent salt as a liquid of 
point-like charges, $\pm e$.  Using the
standard Hubbard-Stratonovich transformation~\cite{ColFluid},
the grand-canonical
partition function $Z$ and free energy $F=- {\beta}^{-1} \log Z [\rho(\rbf)]$ in presence of an external charge distribution
$\rho(\rbf)$ can be written as a functional integral over the field $\phi(\rbf)$ that is conjugate to $\rho(\rbf)$
\bea
\label{e1}
 Z [\rho(\rbf)] &=& \int {\cal D} \phi(\mathbf{r}) \exp \left(-\beta  \int {\rm d}^3\mathbf{r}\ { f}(\phi(\mathbf{r})) \right. \nonumber\\
  & & \left.  -\,i\beta \int {\rm d}^3 \mathbf{r} \, \phi(\mathbf{r}) \rho(\mathbf{r}) \right) 
\eea
with
\bea
\label{e2}
\beta {f}(\phi(\mathbf{r})) &=& \frac{\beta \varepsilon_0}{2}[\nabla\phi(\mathbf{r})]^2
  -\,  2\lambda_s \cos [\beta e \phi(\mathbf{r})] \nonumber\\
& & -\,  \lambda_d \frac{\sin(\beta p_0 |\nabla \phi(\mathbf{r})|)}{\beta p_0 |\nabla \phi(\mathbf{r})|}~,
\eea
where $\beta=1/\kbt$, and $\kbt$ is the thermal energy. The fugacities of the salt $\lambda_s$ and water $\lambda_d$ are determined
by solving the implicit equations
\bea
\label{e3}
\frac{\lambda_s}{V} \frac {\partial}{\partial \lambda_s}  \log Z &=& c_s ~,\nonumber\\
\frac{\lambda_d}{V} \frac {\partial}{\partial \lambda_d}  \log Z &=& c_d ~,
\eea
where $V$ is the volume and $c_s$ and $c_d$ are, respectively, the salt concentration and density
of water molecules.

The electrostatic potential $\Psi(\rbf)$ is given by
\be
\label{e4}
\Psi(\rbf) = -\frac{1}{\beta} \frac{\delta \log Z [\rho(\rbf)] }{ \delta \rho(\rbf)}
= i \langle \phi(\rbf)\rangle_0 ~,
\ee
where the bracket $\langle ...\rangle_0$ denotes the  thermal average
with the Boltzmann weight given in Eq.~(\ref{e1}), for $\rho=0$.

The Gibbs free energy $G$ is a function of $\Psi(\rbf)$ and is defined as the Legendre transform of $F [\rho(\rbf)]$
with respect to $\rho(\rbf)$ through the equations
\bea
\label{e5}
G[\Psi(\rbf)]&=& F [\rho(\rbf)] - \int {\rm d}^3\mathbf{r}\ \Psi (\rbf) \rho (\rbf) ~, \nonumber\\
\Psi(\rbf) &=& \frac {\delta F [\rho(\rbf)]}{\delta \rho(\rbf)} ~,
\eea
from which we deduce the Legendre relation
\bea
\label{e6}
\rho(\rbf) &=& -\frac {\delta G [\Psi(\rbf)]}{\delta \Psi(\rbf)} ~,
\eea

The dielectric tensor $\varepsilon_{\alpha \beta}$ is defined through the Fourier transform:
\be
\label{e7}
\varepsilon _{\alpha \beta} = \frac {\partial ^2}{\partial p_{\alpha} \partial p_{\beta}}
\int {\rm d}^3\mathbf{r}\ {\rm e}^{i \mathbf{p} \cdot \rbf } \frac {\delta ^2 G[\Psi(\rbf)]}
{ \delta \Psi(\rbf) \delta \Psi(0)} \Bigg |_{\Psi =0, \mathbf{p}=0} ~.
\ee
In our present study, the dielectric tensor for isotropic aqueous solutions is diagonal $\varepsilon_{\alpha \beta} =
\varepsilon \delta_{\alpha \beta} $, and our aim
is to calculate its variation as a function of the salt concentration $c_s$.
Following Eq.~(\ref{e7}), we need only to calculate the coefficient of the
$|\nabla \Psi(\rbf)|^2$ terms in the Gibbs free-energy $G[\Psi(\rbf)]$. 

On the mean-field level
the Gibbs free-energy for the Dipolar PB system, $G_{\rm DPB}$,  is
\bea
\label{e7.1}
\beta G_{\rm DPB}[\Psi] &=& -\frac{ \beta\varepsilon_0}{2}(\nabla\Psi)^2
  -\,  {2c_s} \cosh (\beta e \Psi) \nonumber\\
& & -\,\,  {c_d} {g}(u),
\eea
with ${\bf u}= \beta p_0 \nabla \Psi (\rbf)$,
$u=|{\bf u}|$, and ${g}(u) \equiv \sinh u/{u}$.
The Dipolar Poisson-Boltzmann (DPB) equation is an extension of the standard PB equation,
and can be derived via a variational principle
from Eq.~(\ref{e7.1}):
\begin{eqnarray}
\label{DPBeqn}
-\epsilon_0 \nabla^2 \Psi &=&-2c_s e\sinh\left( \beta  e\Psi\right)\nonumber\\
 &+&c_d p_0 \nabla \cdot \left[\frac{\nabla \Psi}
{|\nabla \Psi|} {\cal G}(\beta p_0 |\nabla \Psi|)\right],
\end{eqnarray}
where the function ${\cal G} =g'(u)=\cosh u/u - \sinh u/u^2 $ is related to the
Langevin function $L(u) = \coth(u) - 1/u$.


The dielectric constant $\varepsilon_{_{\rm DPB}}$ is calculated by substituting $G_{\rm DPB}$
into Eq.~(\ref{e7}). The result is the same as in Ref.~\cite{DPB}:
\be
\label{e8}
\varepsilon_{_{\rm DPB}}=\varepsilon_0+\beta p_0^2c_d/3 ~.
\ee
As can be seen above, the dielectric response depends on the dipole (but not the ion) density $c_d$ and, hence,
cannot explain any dielectric decrement.
At room temperature, $T=300\,$K,
and for pure water with dipolar moment $p_0 =1.8$\,D, and density $c_d=55$\,M, the obtained value of $\varepsilon$ 
is $\varepsilon_{_{\rm DPB}}\simeq 11.1$. Note that this value is much smaller than the
measured water value $\varepsilon_w\simeq 80$. This is not surprising since Eq.~(\ref{e8})
is a dilute gas approximation and does not capture
the correlation effects in the case of dense liquid water as well as the finite-size of water molecules.

It is still possible, however, to circumvent this drawback by
solving the  DPB equation, Eq.~(\ref{DPBeqn}), around
a fixed point-like ion, and
showing the existence of a hydration shell around it. 
Expanding $\cal{G}$ in a Taylor series, the dielectric constant is
extracted as a function of the distance $r$ from the ion:
\be
\label{e12}
\epsilon(r)  \simeq  \frac{\epsilon_{_{\rm DPB}}}{3 h^2\left({l_h}/{r}\right) + 1}\, ,
\ee
where the function $h(u)$ is obtained by solving a cubic equation:
\begin{eqnarray}
\label{e13}
h(u)= \left[ \sqrt{\frac{1}{27}+\frac{u}{4}} +\frac{\sqrt{u}}{2}\right]^{1/3} \nonumber\\
- \left[\sqrt{\frac{1}{27}+\frac{u}{4}} - \frac{\sqrt{u}}{2}\right]^{1/3}.
\end{eqnarray}
The length, $l_h^2 = {\BejLen d/\sqrt{10}}$ in Eq.~(\ref{e12}), depends both on the dipole size
$d=p_0/q$ and the Bjerrum length, $\BejLen=e^2/4\pi\varepsilon \kbt$.
It is the only length scale in the problem characterizing the hydration shell thickness as can be seen from the
behavior of $\epsilon(r)$ (inset of Fig.~1). In the vicinity of the ion ($r\le l_h$), the dielectric response is very
small and it smoothly rises to bulk values as the influence of the ion decreases, within distances of a few $l_h$.

By averaging $\varepsilon$ of Eq.~(\ref{e12})
over a sphere of radius $R$ around each ion, and equating $R$ with the typical
distance between two ions, $(2c_s)^{-1/3}/2$, at concentration $c_s$, we obtain the expression
of $\langle\varepsilon(c_s)\rangle$. As can be seen in Fig.~1, the non-linear dielectric
decrement is reproduced and fits rather well the
experimental data for RbCl and CsCl salts of Ref.~\cite{Experiment2}. For low salt concentration,
the averaging of $\varepsilon$ can be done analytically and results in
 $\gamma\simeq 110 \epsilon_{_{\rm DPB}} l_h^3 c_s$, which
is proportional to the hydration shell volume. With parameter values as in Fig.~1, we get $\gamma\simeq 17$.
%
\begin{figure}[ht]
\includegraphics[scale=0.45]{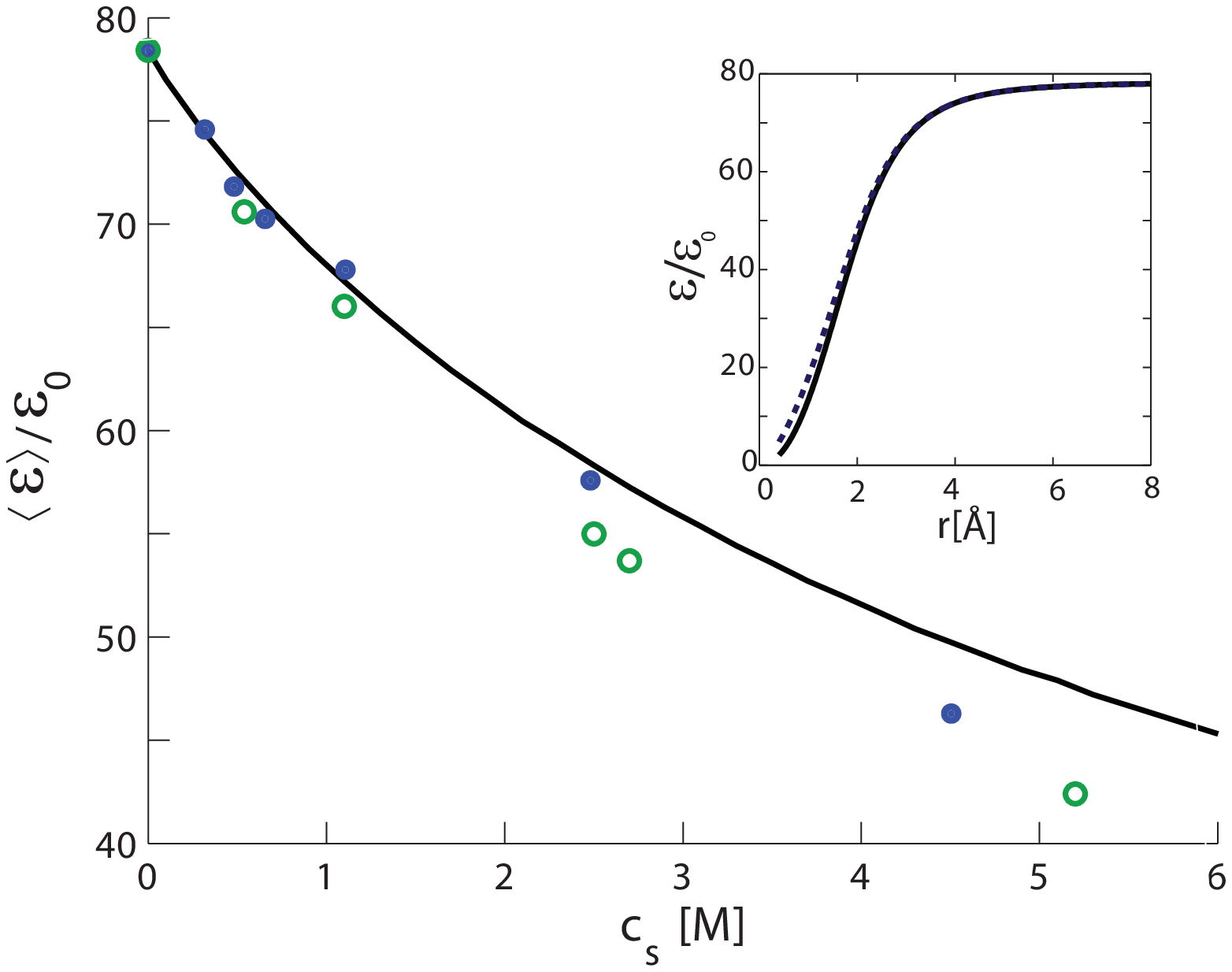}
\caption{\textsf{(color online).  The dielectric constant $\langle\varepsilon\rangle$
averaged inside a specific volume around a single ion (solid line)  as function of ionic concentration, $c_s$.
The comparison is done with experimental values for RbCl (empty circles) and CsCl (full circles)~\cite{Experiment2}.
In the inset, the exact  (solid line) and  approximated  (dashed, Eq.~(\ref{e12})) solutions of the DPB equation, Eq.~(\ref{DPBeqn})
are shown as function of the distance $r$ from a point charge (ion). Choosing as a fit parameter the dipole moment
of water to be $p_0=4.8$\,D (instead of the physical value $p_0=1.8$\,D)~\cite{DPB}, 
allows us to obtain $\varepsilon_{_{\rm DPB}}=80$ and $l_h\simeq 1.5$\,\AA.
}}
\label{fig1}
\end{figure}

So far in order to induce the necessary ion-dipole correlations, we had to rely on calculating the
dielectric response around a single fixed ion, which gives the dielectric
decrement in an approximated way. Moreover, the validity of the DPB theory can be justified only in the dilute limit,
where the dipole-dipole fluctuations are not important.

To overcome this limitation
a more complete treatment of the statistical
mechanics of ionic and dipolar degrees of freedom (including their interactions) is developed.
We generalize the Debye-H\"uckel approximation to
dipolar systems in presence of salt,
via a loop expansion~\cite{LoopExp1,LoopExp2} of the Gibbs free-energy, $G=G_{\rm DPB}+\Delta G$.
To one-loop order the calculation of $\Delta G$ yields:
\bea
\label{e7.2}
\Delta G[\Psi(\rbf)] &=& \frac{1}{2\beta} {\rm Tr} \log\left[-\varepsilon_0
\nabla^2+ 2 \lambda_s \beta e^2 \cosh( \beta  e \Psi(\rbf)) \right.  \nonumber \\
&+& \left. \lambda_d \beta p_0^2\left( {\partial_{\alpha}}\Gamma_{\alpha \gamma}
(\rbf){\partial_{\gamma}}+  \Gamma_{\alpha \gamma}(\rbf) {\partial_{\alpha}}\cdot
{\partial_{\gamma}}\right) \right] ~,
\eea
with
\bea
\label{e7.3}
\Gamma_{\alpha \gamma}(\rbf) =\delta_{\alpha \gamma}
\frac{g'(u)}{u} - \frac{u_{\alpha} u_{\gamma}}{u^2}\left(\frac{g'(u)}{u} -
g''(u) \right ) ~.\nonumber\\
\eea


To get a consistent expression for the dielectric constant we  need to pay extra attention
in the loop expansion to the expressions of the ion and dipole fugacities, $\lambda_s$ and $\lambda_d$.
While at mean-field level $\lambda_s$ and $\lambda_d$
are equal to the ion and dipole (water) densities,
$c_s$ and $c_d$, respectively, their corrections are derived by expanding Eq.~(\ref{e3}) to one-loop order.
To this order, only the correction to $\lambda_d$  will affect
the mean-field value of the dielectric constant, $\varepsilon_{_{\rm DPB}}$,
and is written as
\bea
\label{e9}
\lambda_d  &=&  c_d  -  \frac{2\pi}{3a^3} \frac{\varepsilon_{_{\rm DPB}}-\varepsilon_0}
{ \varepsilon_{_{\rm DPB}}}\left[1  - \,\frac{3}{4\pi^2} ({\DebFreq}a)^2 \right. \nonumber \\
& +&\left. \frac{3}{8\pi^3} ({\DebFreq}a)^3\tan^{-1}\frac{2\pi}{\DebFreq a}\right]~,
\eea
where $a$ is a microscopic cutoff length and $\DebFreq=\sqrt{8\pi\BejLen c_s}$ is the inverse Debye length. The  origin of the
cutoff in our formulation is related to the fact that Coulombic interactions diverge at zero distance,
while in reality such a divergence is avoided because of steric repulsion. A simple way to take
this into account is through a cutoff length $a$, which is related to the minimal distance between adjacent
dipoles and charges, and thus indirectly also to the size of dipoles and ions.

We can now calculate consistently the corrections to the dielectric constant up to one-loop order.
The correction term can be split into water and salt contributions,
$\varepsilon-\varepsilon_{_{\rm DPB}}=\Delta\varepsilon_{d}+\Delta\varepsilon_{s}$:
\bea
\label{e10}
\Delta\varepsilon_{d}&=& \frac{(\varepsilon_{_{\rm DPB}}-\varepsilon_0)^2} {\varepsilon_{_{\rm DPB}}} \frac{4\pi}{3 c_d a^3}\nonumber\\
\Delta\varepsilon_{s}&=&-\frac{(\varepsilon_{_{\rm DPB}}-\varepsilon_0)^2 } {\varepsilon_{_{\rm DPB}}} \frac{\DebFreq^2}{\pi c_d a}\left( 1 -
\frac{\DebFreq a}{2\pi}\tan^{-1}\frac{2\pi}{\DebFreq a}\right) \nonumber\\
\eea
The term $\Delta\varepsilon_{d}$ represents the fluctuation effect  of the water dipoles beyond
the mean-field DPB level. It varies as $\sim 1/(c_d a^{3})$.
This pure water fluctuation term essentially adds a positive numerical prefactor of rather
large magnitude to the mean-field value of $\varepsilon_{_{\rm DPB}}$,
meaning that the one-loop correction is important even for the pure water case.

The second term $\Delta\varepsilon_{s}$ has by itself two contributions. The leading term in the dilute
solution limit, $\DebFreq a\ll 1$,   depends linearly on the salt concentration, $\Delta\varepsilon_{s}=-\gamma c_s$, with
\be
\label{e10.1}
\gamma=\frac{(\varepsilon_{_{\rm DPB}}-\varepsilon_0)^2 } {\varepsilon_{_{\rm DPB}}} \frac{8\BejLen}{ c_d a} ~.
\ee
When the Debye length $\DebFreq^{-1}$ is of the
same order of magnitude as $a$ (high salt limit), the last term in $\Delta\varepsilon_{s}$  starts to dominate and the
dielectric decrement becomes smaller until eventually it will reverse the trend and may even cause a {\it dielectric
increment} as seen in experiments~\cite{Experiment2} in the high salt limit.

We compare our prediction for the dielectric constant $\varepsilon$, Eq.~(\ref{e10}), to experimental values~\cite{Experiment2}
for seven different ionic solutions in concentration range of 0--6\,M.
We separate the seven salts into three subgroup according to the
size of the alkaline cations as presented in the three parts of Fig.~2. In each of the figure parts the $a$
parameter is fitted separately. We treat $a$
as a free parameter and find its value by the best fit of our prediction, Eq.~(\ref{e10}), to experimental data,
while keeping the physical value of the water dipolar moment, $p_0=1.8$\,D.
The best fit to the data
can be seen in 2(a) and corresponds to the largest ionic size of Cs$^+$ and Rb$^+$. In (b)
the fit for K$^+$ ions is also quite good, while in 2(c) for the smallest ions, Na$^+$ and Li$^+$,
the deviation is more pronounced, especially for the
larger values of $c_s\ge 4$\,M.

Note that our formula takes into account only in a broad sense the finite size of ions (and the distance of closest approach between them)
via the $a$ parameter. It is beyond the level of the theory to give more specific ionic predictions.
Hence, the obtained value of  $a\simeq 2.7$\,\AA\  is not very sensitive to the type of salt, but
its main contribution comes from the water dipoles themselves whose
diameter~\cite{WaterRadii} is about $2.75$\,\AA.
On the other hand, as can be clearly seen from Fig.~2, important cooperative effects
of ions and dipoles are accounted for in our non-linear expression for $\varepsilon(c_s)$.
For small $c_s$, the dashed line represents the best linear fit and
works well only when $c_s\le 1$\,M, while the non-linear prediction of
Eq.~(\ref{e10}) succeeds in fitting the large concentration range as well~\cite{ft24}.

%
\begin{figure*}[ht]
\includegraphics[scale=0.75,draft=false]{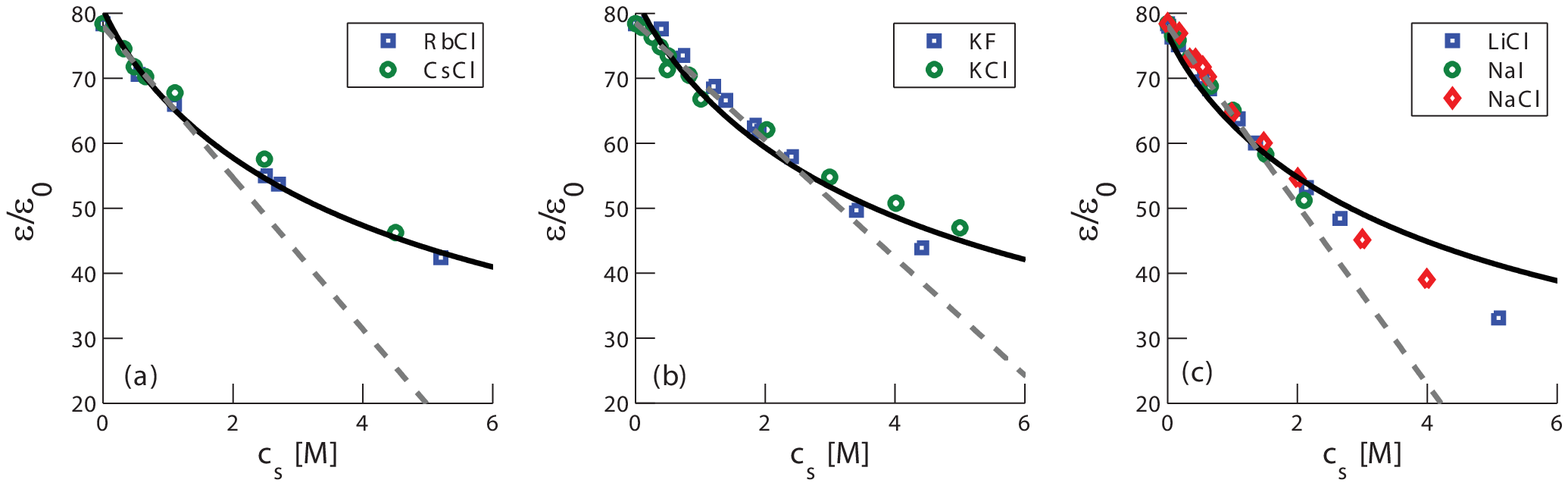}
\caption{\textsf{(color online). Comparison of the predicted dielectric constant, $\varepsilon$, from Eq.~(\ref{e10})
with experimental data from Ref.~\cite{Experiment2}, as function of ionic concentration, $c_s$, for various salts.
The theoretical prediction (solid line) was calculated
using $a$ as a fitting parameter. In (a) the fit is for RbCl and CsCl salts with
$a=2.66$\,\AA; in (b) the fit is for KF and KCl  with $a=2.64$\,\AA; while in (c) the fit is for LiCl,
NaI and NaCl with $a=2.71$\,\AA. The dashed lines are the linear fit to the data in the low
$c_s\le 1$\,M range. The slope of the linear fit is $\gamma=11.7$\,M$^{-1}$ in (a), 9.0\,M$^{-1}$ in (b) and 13.8\,M$^{-1}$ in (c). The value of $\gamma$ for each salt
varies by about 10-20\,$\%$ and the linear fit should be taken as representative for the low $c_s$ behavior.}}
\label{fig2}
\end{figure*}

In conclusion, we are able to reproduce rather well the linear and non-linear dielectric decrement behavior
over a large range of ionic
concentrations, and the obtained values of $\varepsilon$ are in quantitative agreement with the
data for several types of monovalent salts. In addition, we found a qualitative
description of the hydration shell characterized by a single length scale, $l_h$.
Note that our model does not
contain any significant ionic-specific effects. To improve on the latter, we would need to
include the ionic finite-size and specific non-electrostatic short-range interactions
that affect both bulk properties of ionic solutions such as dielectric constant and viscosity
as well as their behavior at interfaces and, in particular, their surface tension.

{\bf Acknowledgements.~~~} {We thank D. Ben-Yaakov, Y. Burak and X.-K. Man for useful discussions. This work
was supported in part by the U.S.-Israel Binational Science Foundation under Grant No.
2006/055 and the Israel Science Foundation under Grant No. 231/08. One of us (HO)
would like to thank  the Raymond \& Beverly Sackler Program for Senior Professors by Special Appointment at Tel Aviv University.}



\begin{thebibliography}{99}
\bibitem{ft24} In the low $c_s$ limit, we estimate $\gamma\simeq26$ from Eq.~(\ref{e10.1}) and for $a=2.7$\,\AA.
\end{thebibliography}
\end{document}